\documentclass[prb, aps, twocolumn, groupedaddress,showpacs]{revtex4}
\usepackage{eqnarray}
\usepackage[english]{babel} 
\usepackage[english]{layout}
\usepackage{amsmath,amsfonts,amssymb}
\usepackage{graphicx}
\usepackage{float}
\usepackage{color}
\usepackage{braket}
\usepackage{version}
\usepackage{array,multirow,makecell}

\begin{document}

\title{Interacting quantum fluid in a polariton topological insulator}
\author{O.Bleu, D. D. Solnyshkov, G. Malpuech}
\affiliation{Institut Pascal, PHOTON-N2, Clermont Universit\'{e}, Blaise Pascal
University, CNRS, 24 avenue des Landais, 63177 Aubi\`{e}re Cedex, France.}

\begin{abstract}

We study Bogoliubov excitations of a spinor Bose Einstein condensate in a honeycomb periodic potential, in the presence of a Zeeman field and of a spin-orbit coupling specific for photonic systems, which is due to the energy splitting between TE and TM polarized eigenstates. We also consider spin-anisotropic interactions typical for cavity polaritons.  We show that the non-trivial topology of the single particle case is also present for the interacting system. At low condensate density, the topology of the single-particle bands is transferred to the bogolon dispersion. At a critical value, the self-induced Zeeman field at the Dirac points of the dispersion becomes equal to the real Zeeman field and then exceeds it. The gap is thus closed and then re-opened with inverted Chern numbers. This change of topology is accompanied by a change of the propagation directions of the one-way edge modes. This result demonstrates that the chirality of a topological insulator can be reversed by collective effects in a Bose-Einstein condensate.

\end{abstract}

\maketitle

The concept of topological insulators relies on the chirality of the Bloch bands. This concept was first introduced for electronic systems \cite{thouless1982quantized,haldane1988model,kane2005z,bernevig2006quantum,hasan2010colloquium} and then extended to photonic \cite{haldane2008possible,wang2008reflection,Soljacic2009,Soljacic2014}, and more generally to bosonic systems \cite{aidelsburger2015measuring,peano2014topological}. In this last case, the non-trivial topology of the system is addressed by either resonant excitation in photonics, or by driving an atomic gas strongly out of equilibrium \cite{jotzu2014experimental}. Another field of research, extremely fruitful, is the one dealing with collective effects. In fermionic systems, the combination of non-trivial topology and many body effects led to the most intriguing phenomena in physics such as the fractional quantum Hall effect \cite{laughlin1983anomalous}, topological superconductors \cite{Fu2008,hasan2010colloquium}, and a vast variety of other phenomena \cite{wang2014classification}. In bosonic systems, the most well-known collective effect is the Bose-Einstein Condensation \cite{Einstein,lnoir}, leading to fascinating dynamical behaviour such as superfluidity \cite{RevModPhys.71.S318}, and, more generally, to the physics of quantum fluids \cite{VolovikReview,Leggettbook,carusotto2013quantum}. So far, the physics of bosonic quantum fluids in topologically non-trivial systems has been addressed only in a couple of works, essentially dealing with atomic condensates \cite{engelhardt2015topological,bardyn2015chiral,furukawa2015excitation}. 

On the other hand, the mixed exciton-photon quasiparticles called exciton-polaritons represent a very good implementation of quantum fluids of light \cite{carusotto2013quantum} in which a wide variety of phenomena such as polariton BEC \cite{Kasprzak2006}, superfluidity \cite{Amo2009}, quantized vortices \cite{Lagoudakis2009} have been observed. 
Polaritons possess the same type of spin-orbit coupling as other photonic systems, based on energy splitting between TE and TM polarized eigenmodes. (For a review on polariton spin effects see Ref.\cite{Shelykh2010}). The specific relation between the polarization of the TE and TM modes and their  propagation direction induces an intrinsic chirality for the photonic system which serves as a basis for several effects, such as the Hall effect for light \cite{Onoda2004} and optical spin Hall effect \cite{Kavokin2005,Leyder2007}. When this spin-orbit coupling (SOC) is combined with a finite Zeeman field, the propagation along a curved trajectory leads to the appearance of a non-zero Berry phase\cite{Shelykh2009,KarzigPRX2015}. In a lattice with band crossing at the edge of the Brillouin zone, for instance in honeycomb lattices, the Berry phase accumulates along the band, and the corresponding Chern number is different from zero. This configuration is the one initially proposed by Haldane and Raghu to achieve topologically non-trivial photonic bands  \cite{haldane2008possible} and one-way edge states. This anomalous quantum Hall effect   \cite{haldane1988model} for photons has been experimentally evidenced soon after its prediction, the Zeeman field being produced by a finite gyromagnetic response of the material used in the microwave range of wavelengths. Similar ingredients have been recently used to demonstrate the feasibility of a topological insulator analog for polaritons \cite{nalitov2015spin,nalitov2015polariton,LiewPRB2015, KarzigARXIV2015}. In this last case, the Zeeman field is linked with the exciton g-factor, which should allow the implementation of one-way edge states at optical frequencies. Another key advantage of using mixed light-matter states is, as previously mentioned, their interacting character, which allows to study the behavior of an interacting quantum fluid in a topologically non-trivial band structure.

In this work we consider bosons with two spin components and spin-anisotropic interactions. The gas of bosons is assumed to be at thermal equilibrium in a honeycomb lattice. We consider TE-TM induced SOC and a Zeeman splitting created by an external magnetic field. This configuration is typical for cavity polaritons, and has the advantage to allow the existence of a stable homogeneous condensate in the ground state, which is not the case for other types of Rashba/Dresselhauss SOC \cite{lin2011spin}, available in atomic systems. 
In the first section, we remind the tight binding description of the polariton graphene in the linear regime, including photonic SOC and Zeeman splitting.  The band structure \cite{nalitov2015spin} is characterized by a non-zero topological invariant \cite{nalitov2015polariton} which defines a $\mathbb{Z}$ topological insulator. In the second section, we consider a polariton condensate in this system, at thermal equilibrium at $T=0$ K in presence of a magnetic field. The condensate and its weak excitations are described by a spinor Gross-Pitaevskii equation. At low density, the equilibrium condensate is circularly polarized, because the real Zeeman splitting exceeds the self-induced Zeeman splitting, the latter being different from zero because of the spin-anisotropic polariton-polariton interactions \cite{Renucci2005, Shelykh2010}. We find that the topology of the single particle band transfers to the bogolon dispersion, similar to the conclusions of Ref. \cite{furukawa2015excitation}. Both polariton and bogolon bands are characterized by the same non-zero topological invariants. The density rise leads to an increase of the compensating self-induced Zeeman field, bringing closer the $\sigma^+$ and $\sigma^-$ bands. We find that the self-induced Zeeman field is approximately twice larger at the Dirac point than at $k=0$. As a result, the gap between the bands closes first at the Dirac points of the dispersion for a threshold density $n _S =  \mu g_X B/2(\alpha_1-\alpha_2)$, where $\alpha_{1,2}$ are the interaction constants in the singlet and triplet configurations, respectively \cite{vladimirova2010polariton}. Increasing the condensate density immediately re-opens the gap with a reversed sign of the band Chern numbers. This topology inversion is associated with an inversion of the propagation direction of the edge modes. Increasing  the density further suppresses the splitting at $k=0$, this time preserving the topology of the bands, and the condensate becomes elliptically polarized (spin-Meissner effect) \cite{Rubo2006}.

\section{Polariton Z topological insulator}

Cavity polaritons are composite particles appearing in the regime of strong light-matter (photon-exciton) coupling in optical microcavities \cite{PhysRevLett.69.3314,kavokin2003cavity}. These are photonic states but with an exciton part, which allows them to interact with each other, a feature at the heart of the quantum fluid behaviour of polariton gas. 
The heavy-hole excitons are characterized by angular momentum projections equal to $\pm 1$ or $\pm 2$. Due to the optical selection rules, only excitons with spin $\pm 1$ can be coupled with light. One can therefore create polaritons with circularly polarized photons ($\sigma^+ ,~\sigma^-$).
Hence, the spin structure of a polariton is similar to the one of electron (both are two-level systems) and the condensate wave function is a linear combination of the two polarizations, that is, a spinor - a vector with 2 complex components:

\begin{equation}
\mathbf{\Psi}(\textbf{r},t)=\begin{pmatrix}
\Psi^+(\textbf{r},t) \\ \Psi^-(\textbf{r},t)
\end{pmatrix}
\end{equation}

In photonic structures, the eigenmodes are TE and TM polarized. The Hamiltonian acting on the polariton wave function in the basis of the circular-polarized reciprocal space states with wavevector $\mathbf{k}$ reads:
\begin{equation}
\hat{H}=\begin{pmatrix}
\frac{\hbar^2 k^2}{2m}&-\frac{\Omega_{TE-TM}}{2}e^{-2i\phi}\\
-\frac{\Omega_{TE-TM}}{2}e^{2i\phi}&\frac{\hbar^2 k^2}{2m}
\end{pmatrix}
\end{equation}
where $\phi$ is the angle between the wavevector $\mathbf{k}$ and the horizontal axis, and the value of the TE-TM splitting $\Omega_{TE-TM}$ depends on the wavevector magnitude $k$. This Hamiltonian can be expressed as a superposition of the kinetic energy term and a linear superposition of Pauly matrices :

\begin{equation}
\hat{H}=\frac{\hbar^2 k^2}{2m}\mathbf{I}-\mathbf{\Omega}.\boldsymbol{\sigma}
\end{equation}

This decomposition allows to define the effective $k$-dependent magnetic field $\mathbf{\Omega}$ acting on the polariton pseudo spin, which is the photonic/polaritonic spin-orbit coupling. This effective field has two non-zero components and a winding number 2. In a planar cavity, it is in the plane and its magnitude is approximately proportional to $k^2$ close to $\mathbf{k}=0$:
\begin{equation}
\mathbf{\Omega}=\beta k^2(\cos 2\phi, \sin 2\phi,0)
\end{equation}

The effective field texture around $k=0$ in a microcavity is shown on Fig. 1(a). It demonstrates a clear dipolar structure. Figure 1(b) shows the effective field texture at a given energy. The double winding of the effective field can be clearly visualized.
\begin{figure}[H]
 \begin{center}
 \includegraphics[scale=0.4]{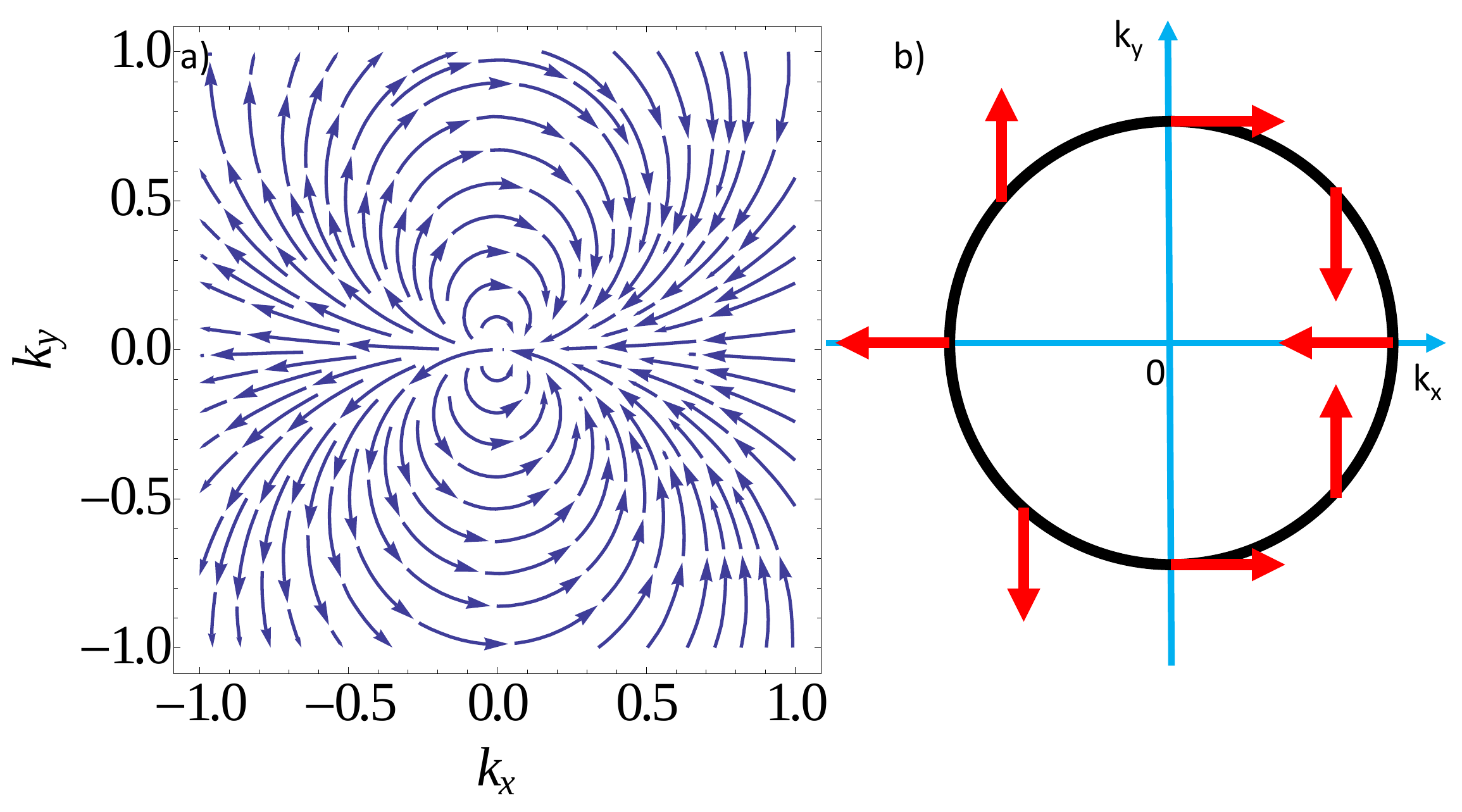}
 \caption{ (Color online) (a) TE-TM field pattern in the reciprocal space (b) TE-TM field for an elastic circle (fixed energy).}
  \end{center}
 \end{figure}
 
The lower energy state corresponds to a spinor aligned with the effective field and the corresponding wave function reads:
\begin{equation}
\psi=\psi_k\begin{pmatrix}
e^{-i\phi}\\e^{i\phi}
\end{pmatrix}
\end{equation}


During the last years, experimental groups have fabricated a lot of complex samples with different periodic in plane energy potentials by different techniques, such as metal deposition \cite{kim2013exciton}, the use of surface acoustic waves \cite{PhysRevLett.105.116402}, of mesas \cite{Paraiso2010,Hofling2015}, or etching planar microcavities. This last techniques allows to fabricate coupled polariton micropillars \cite{microp} organized in molecules \cite{sala2015spin} or   lattices \cite{jacqmin2014direct}.\\

\begin{figure}[H]
 \begin{center}
 \includegraphics[scale=0.41]{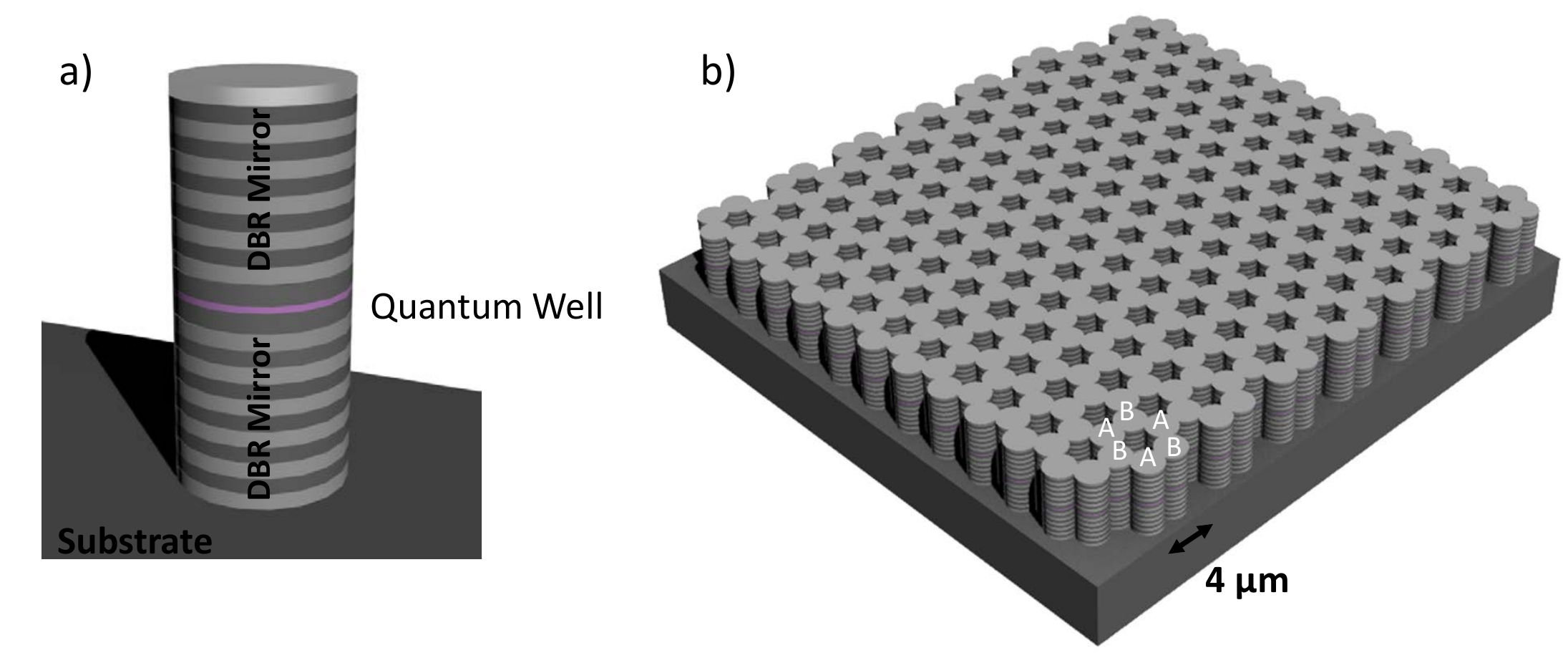}
 \caption{ (Color online) (a) Polaritonic micropillar scheme, (b) polaritonic graphene scheme.}
  \label{scheme}
  \end{center}
 \end{figure}

We therefore consider a honeycomb lattice of polariton micropillars (Fig. \ref{scheme}) \cite{jacqmin2014direct}. As in conventional graphene, we can define two triangular sub-lattices A and B. Then, the system acquires two pseudospins: one associated with the polarization, as above, and the other corresponding to the sublattice degree of freedom (A/B).
Taking into account the TE-TM SOC \cite{nalitov2015spin}, tunneling coefficients between nearest neighbors are expressed on the circular basis as: 
\begin{eqnarray}
 \bra{A,\pm}H\ket{B,\pm}&=&-J  \\ \bra{A,\pm}H\ket{B,\mp}&=&-\delta J e^{-2i\phi_j} \nonumber
 \end{eqnarray}
 $\phi_j$ is angle of $\textbf{d}_{\phi_j}$ vector with horizontal direction. $J$ is the tunneling  coefficient without spin inversion, like in conventional graphene. $\delta J$ is the spin orbit coupling term due to TE-TM splitting, $\delta J=(J_L-J_T)/2$, where $J_L$ and $J_T$ are the tunneling coefficients for the longitudinally and transversally-polarized polaritons respectively. 
 
 A polaritonic state will be written as a bispinor 
 (A-B sub-lattices and $\pm$ pseudo-spin): $\Phi=(\Psi_A^+,\Psi_A^-,\Psi_B^+,\Psi_B^-)^T$.
The effective Hamiltonian written in tight binding approximation has a unitary block matrix form:  
\begin{eqnarray}
H_k=\begin{pmatrix}
0&F_k \\
F_k^+&0
\end{pmatrix} 
,\quad F_k=-\begin{pmatrix}
f_kJ&f_k^+\delta J \\
f_k^- \delta J&f_kJ
\end{pmatrix}
\end{eqnarray}
where the complex coefficients $f_k$ and 
$f_k^{\pm}$ are defined by:
\begin{eqnarray}
f_k&=&\sum_{j=1}^3\exp{(-i\textbf{kd}_{\phi_j})} \\ f_k^{\pm}&=&\sum_{j=1}^3 \exp{(-i[\textbf{kd}_{\phi_j}\mp 2\phi_j])}\nonumber
\label{fk}
\end{eqnarray}
This Hamiltonian can be diagonalized analytically \cite{nalitov2015spin} and degeneracies occur at the edges of the Brillouin zone -- the K and K' points. \\

We can linearize the Hamiltonian close to the Dirac points ($\textbf{q}=\textbf{k}-\textbf{K}$, $q\ll \pi/a$). It contains two distinct contributions:
\begin{equation}
H=H_0+H_{SO},
\end{equation}
where $H_0$ is the usual Dirac Hamiltonian for a massless particle, similar to graphene:
\begin{equation}
\begin{array}{l}
H_0=\hbar v_f (\tau_z q_x \sigma_x +q_y \sigma_y)
\end{array}
\end{equation}
The spin-orbit coupling part reads:
\begin{equation}
H_{SO}=3\delta J/2\left(\tau_z \sigma_y s_y - \sigma_x s_x\right)
\end{equation}
In the above expressions, $\hbar v_F=3Ja/2$ and $\Delta=3\delta J/2$. If the SOC is relatively weak ($\delta J/J\ll 1$), it is possible to consider a region of wavevectors not exactly close to $q=0$, but at the same time sufficiently small, where $\delta J/J \ll qa \ll 1$. In this region, the spin orbit term plays a role of perturbation over the Dirac Hamiltonian $H_0$ and may thus be rewritten as an interaction with an in-plane $q$-dependent magnetic field:

\begin{equation}
H_{SO}=-\gamma \tau_z( q_x s_x -q_y s_y)/q
\end{equation} 
where 
$v_f=3Ja/(2\hbar)$, $\gamma=3\delta J/2$, and $\tau_z$ equals $+1$ at K or $-1$ at K'. $\hat{\sigma}$ and $\hat{s}$ are spin operators acting on A/B sublattice and polarization degrees of freedom respectively. In this case, the two pseudospins are decoupled, and one can definitely speak of an effective magnetic field acting on the polarization pseudospin only. 

However, close to the Dirac point where $q=0$ or in the case when $\delta J/J>0.5$, such approximation is not possible. Still, even in this case it is possible to separate the polarization and lattice pseudospins by partial diagonalization of the Hamiltonian in the basis of bonding and antibonding states with linear polarization corresponding to the texture of the Dresselhaus field, e.g. ${\psi _1} = {\left( { - 1, - {e^{ - i\phi }},{e^{i\phi }},1} \right)^T}$.  After the separation of the effective field acting only on the polarization pseudospin, it can be converted back to the original basis, demonstrating an in-plane texture identical to that of the Dresselhaus field:
\begin{equation}
H'_{SO}=q_x s_x - q_y s_y
\end{equation}
with an important contribution of a field in the $Z$ direction, proportional to $\Delta$ and opposite on A and B atoms. It is this $q$-independent field, which induces a parabolic dispersion at the Dirac points. One should note that the lattice part of the Hamiltonian is not $H_0$ anymore.

\begin{figure}[H]
 \begin{center}
 \includegraphics[scale=0.5]{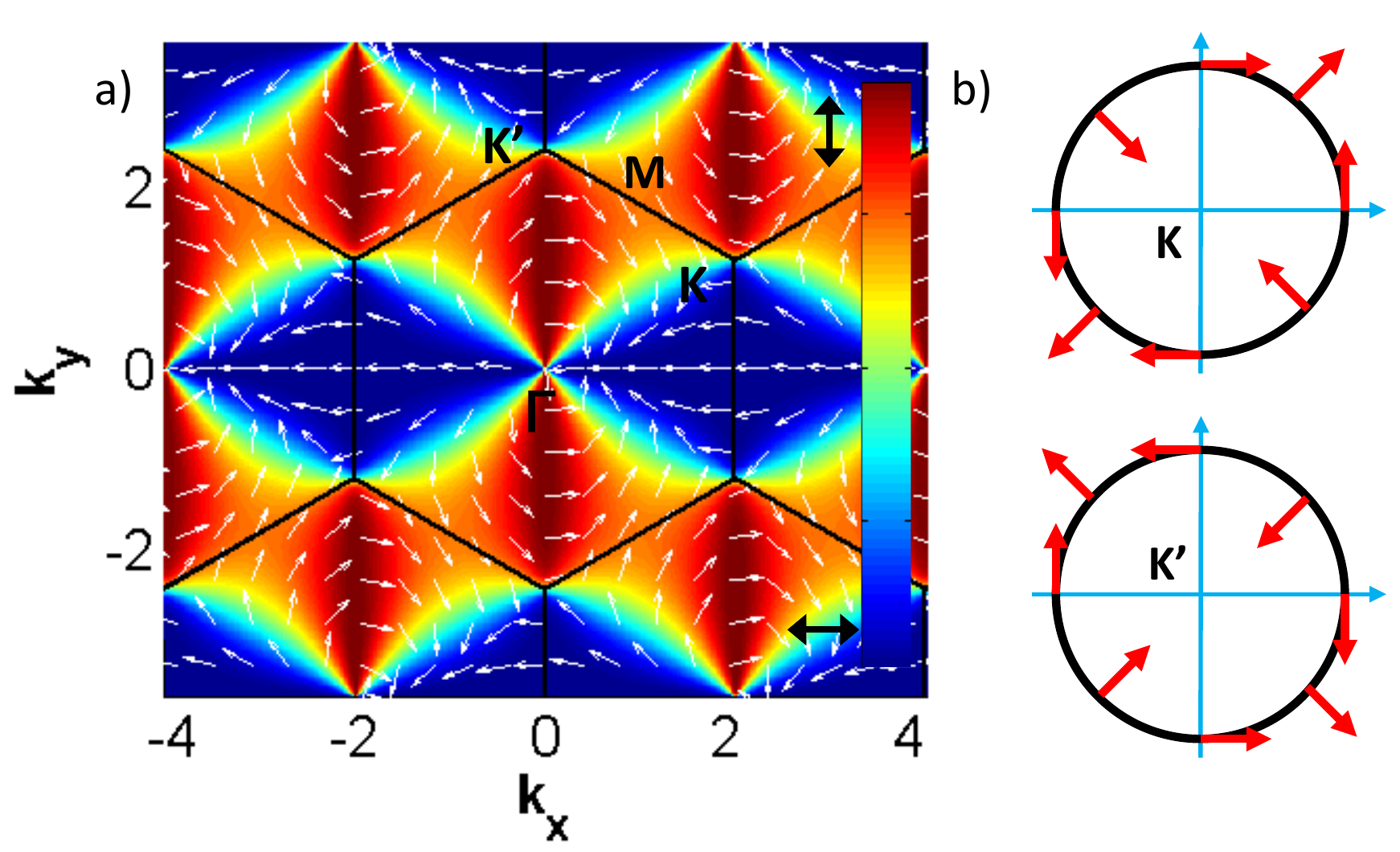}
 \caption{ (Color online) (a) Texture of TE-TM effective field in reciprocal space (white arrows), the colors show the corresponding in-plane polarisation domains, (b) TE-TM Dresselhaus-like field for an elastic circle at K and K' points.}
  \label{dresselhaus}
  \end{center}
 \end{figure}

In order to assess qualitatively the topology of the bands, we can look at the textures of the two pseudospins close to the Dirac point, where a gap opens under applied magnetic field. The effective field texture acting on the sub-lattice pseudospin (Hamiltonian $H_0$) is the one of the standard Dirac Hamiltonian, providing opposite winding of the corresponding spinor at the K and K' points. The texture of the effective magnetic field induced by the TE-TM splitting ($H_{SO}$) in the reciprocal space is shown in Fig. \ref{dresselhaus}. This figure confirms that, according to the linearization above, the texture of the pseudo spin is the one of a Dresselhauss field. The sign of the field changes between K and K', but not the direction of its winding, which allows the Berry phase to add up and give a nonzero topological invariant.
The resulting polaritonic states  are completely projected on $\Psi_A^-$ and $\Psi_B^+$ at K point, and on $\Psi_A^+$ and $\Psi_B^-$ at K'.

 One can open a trivial band gap by breaking spatial inversion symmetry, for instance, by making A and B atoms different. Indeed, the Berry phase accumulated at K and K' is opposite for the sublattice pseudospin, because of the opposite winding of the effective field. On the contrary, a band gap induced by the breaking of the time reversal symmetry, for instance, by applying a finite Zeeman field leads to the formation of non trivial bands with non-zero Chern numbers. Indeed, in this last case, the Berry phase accumulated near K and K' has the same sign, because of the same winding of the effective field acting on the polarisation pseudo-spin. One should notice that this picture is fully similar with respect to the one described by Haldane and Raghu in 2008 \cite{haldane2008possible}. They proposed to use a photonic crystal waveguide which allows to consider a large energy differences between TE and TM modes. However, when considering polariton graphene based on a lattice of coupled pillars, the large TE-TM splitting condition is typically not fulfilled ($\delta J/J=0.1<0.5$). In this case, principally addressed in \cite{nalitov2015spin,nalitov2015polariton,LiewPRB2015},  a quadratic spin-orbit coupling term has to be taken into account in the Hamiltonian close to the Dirac point, leading to the formation of four Dirac points in each valley (trigonal warping effect). 
 
 To analyze the topology of the bands quantitatively, we use a gauge-invariant method based on the Berry curvature:
\begin{equation}
C_n=\frac{1}{2i\pi}\int_{BZ} \mathbf{F}_n(\mathbf{k}) d\textbf{k}
\label{chern}  
\end{equation}
where the Berry connection $\mathbf{A}$ and the associated Berry curvature $\mathbf{F}$ are given by
\begin{eqnarray}
\mathbf{A}(\mathbf{k})&=&\bra{\Phi_{\textbf{k},n}}\nabla_{\textbf{k}} \ket{\Phi_{\textbf{k},n}}\\
\mathbf{F}_n(\mathbf{k})&=&\nabla_{\textbf{k}}\times\mathbf{A}
\end{eqnarray}

This approach allows to calculate the Chern numbers in a very efficient way \cite{fukui2005chern}.
We define two Chern invariants. Each of them is the sum of the Chern numbers of the two branches forming "valence" and "conduction" bands:
\begin{equation}
C_{V}=C_{V_1}+C_{V_2}, ~~ C_{C}=C_{C_1}+C_{C_2}
\end{equation}

A 2D map of the Berry curvatures of each of the subbands close to K and the resulting Chern numbers are shown on Fig. 4. One can observe two distinct phases. When $\delta J/J<0.5$ (Fig. 4(a),(b)), the system exhibits trigonal warping with 4 Dirac points for each valley (K or K'). One can see that one of the Dirac points contributes negatively (the central one), whereas the three others contribute positively to the Berry phase. The contribution is the same at K', and therefore the total band Chern numbers take the values $\pm2$, leading to the existence of two chiral states on each edge. When $\delta J/J>0.5$ (Fig. 4(c),(d)), the trigonal warping is absent, and a single Dirac point contributes to each of the K/K' valleys, leading to  Chern numbers $\mp1$. One can notice that at the topological transition $\delta J/J=0.5$ the Chern number of a given band changes not only its value, but also its sign, leading to a change of the direction of propagation for the edge modes.

\begin{figure}[H]
 \begin{center}
 \includegraphics[scale=0.43]{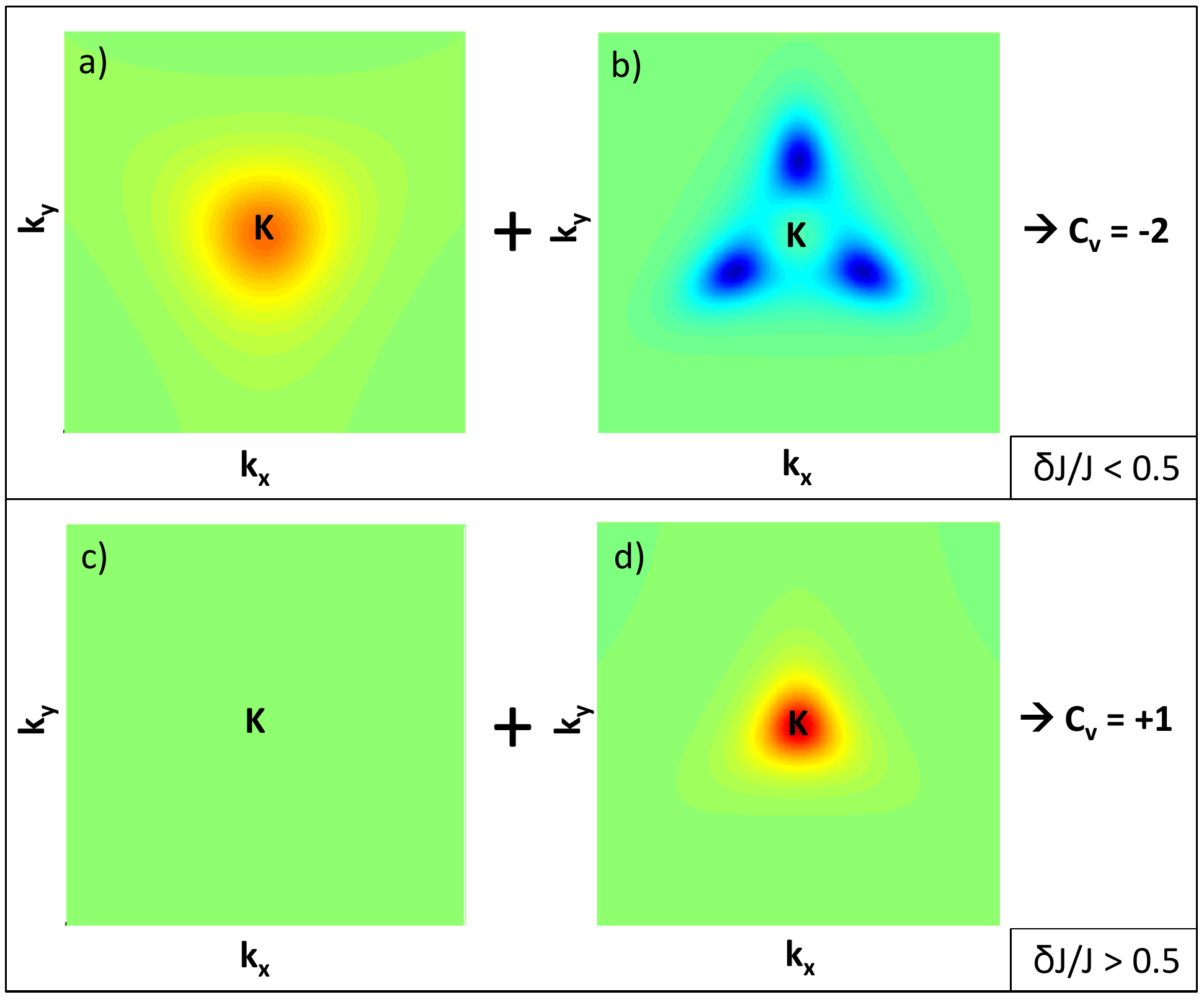}
 \caption{ (Color online) Berry curvatures around K. (a)-(b) Berry curvatures of the first and second "valence" branches respectively for $\delta J/J=0.2<0.5$.  (c)-(d) Similar images for $\delta J/J=0.7>0.5$. Zeeman field is $\Delta=0.1 J$ in both cases.}
  \label{berry}
  \end{center}
 \end{figure}

In the following sections, we study this system in the nonlinear regime and analyze its behavior versus the particule density and the applied magnetic field.

\section{Bogoliubov excitations in polaritonic graphene}

 Cavity polaritons have been well studied in different semiconductor materials and despite their finite and relatively short lifetime, their Bose-Einstein condensation close to thermal equilibrium has been observed in several works \cite{Condensat,PhysRevLett.101.136409, PhysRevB.81.125305, wertz2009spontaneous} dealing with 2D cavities. One should notice, however, than in lattices the formation of gap solitons much more favored than the formation of a homogeneous condensate  in the ground state\cite{Jacqmin2014,Tanese2013} because of the higher overlap of the negative-mass states and a longer lifetime of states with antisymetric wave functions  \cite{Rubo2015}.   Similarly, the slow spin relaxation in the excitonic reservoir makes it difficult to achieve equilibrium between the two spin sub-systems, and prevents the observation of the predicted spin-Meissner effect  \cite{PhysRevB.91.155130}. These two aspects of real systems make the situation that we consider, namely a quasi-thermal equilibrium for spinor polaritons in a periodic lattice, quite unrealistic at the present stage of technology, but of a theoretical interest. However, similar conclusions can be achieved for quasi-resonant pumping, which will be the subject of future works, corresponding to the optimal configuration for the experimental verification of our predictions.

We therefore consider that the polariton graphene is filled by a finite density of interacting polaritons at thermal equilibrium at 0K. The polaritons form a Bose-Einstein condensate which can be described within the mean field approximation by the spinor Gross-Pitaevskii equation:
\begin{equation}
i\hbar\frac{\partial}{\partial t}\Psi_\gamma^\sigma=[H_\gamma^\sigma(\textbf{k})+(\alpha_1|\Psi_\gamma^\sigma|^2+\alpha_2|\Psi_\gamma^{-\sigma}|^2)-\mu]\Psi_\gamma^\sigma
\label{GP-graphene}
\end{equation}
where $\gamma$ is sublattice index and $\sigma=+,-$ is the spin projection. $H_\gamma^\sigma$ is the polaritonic graphene Hamiltonian defined above. $\alpha_1$ and $\alpha_2$ are the interaction constants between particles with same spins and with opposite spins, respectively. These interaction constants in polaritonic systems are in general different \cite{vladimirova2010polariton}, which means that the interactions are spin-anisotropic. The reason is that the exchange in the singlet configuration passes through dark excitonic states. We are going to look for the solution of this equation including the condensate and its weak excitations can be written as a 4-component spinor, taking into account sublattice and spin degrees of freedom (eq.\ref{ansatz}):

\begin{widetext}
\begin{eqnarray}
\mathbf{\Phi}(\textbf{r},t)&=&(\Psi_A^+,  \Psi_A^-, \Psi_B^+, \Psi_B^-
)^T \nonumber \\ &=&\sqrt{n}\begin{pmatrix}\textbf{e}\\ \textbf{e}  
\end{pmatrix}+\begin{pmatrix}\textbf{u}_A  \\ \textbf{u}_B
\end{pmatrix}e^{i(\textbf{k.r}-\omega t)}+\begin{pmatrix}\textbf{v}_A\\ \textbf{v}_B
\end{pmatrix}^*e^{-i(\textbf{k.r}-\omega t)}
 \label{ansatz} 
\end{eqnarray}
\end{widetext}

The first term is the stationary part of the function. It holds the information about condensate's polarization $\textbf{e}$ for the two sublattices (A-B). So, $\textbf{e}$ is a 2-dimensional spinor. For generality, $\textbf{u}_{A(B)},\textbf{v}_{A(B)}$ are also spinors of the form ($\textbf{u}_{A}^+,\textbf{u}_{A}^-$).
Here, the sublattice indexes are defined by A and B, while $\pm$ defines the polariton spin. Finally, $\mathbf{u}$ and $\mathbf{v}$ represent the Bogoliubov complex coefficients. 

By introducing this solution in the GP-equation (eq.\ref{GP-graphene}), and keeping only linear terms in $\mathbf{u}$ and $\mathbf{v}$ (Bogoliubov approximation, see method of ref \cite{lifshitz1980statistical}, problem \S 30), we can first deduce the chemical potential.  The energy dispersion and the eigenstates of the elementary low-amplitude excitations of the condensate -- the so-called bogolons -- can then be found by diagonalizing an $8\times 8$ matrix. An example of such calculation, corresponding to the large magnetic field case (which we will define below) is given in the Appendix of the manuscript. In the following, we calculate and analyze the dispersions and the corresponding eigenstates in different cases: first without a magnetic field, and then under weak and strong fields.

\subsection{Zero magnetic field.}

\begin{widetext} 

 \begin{figure}[H]
 \centering
 \includegraphics[scale=0.72]{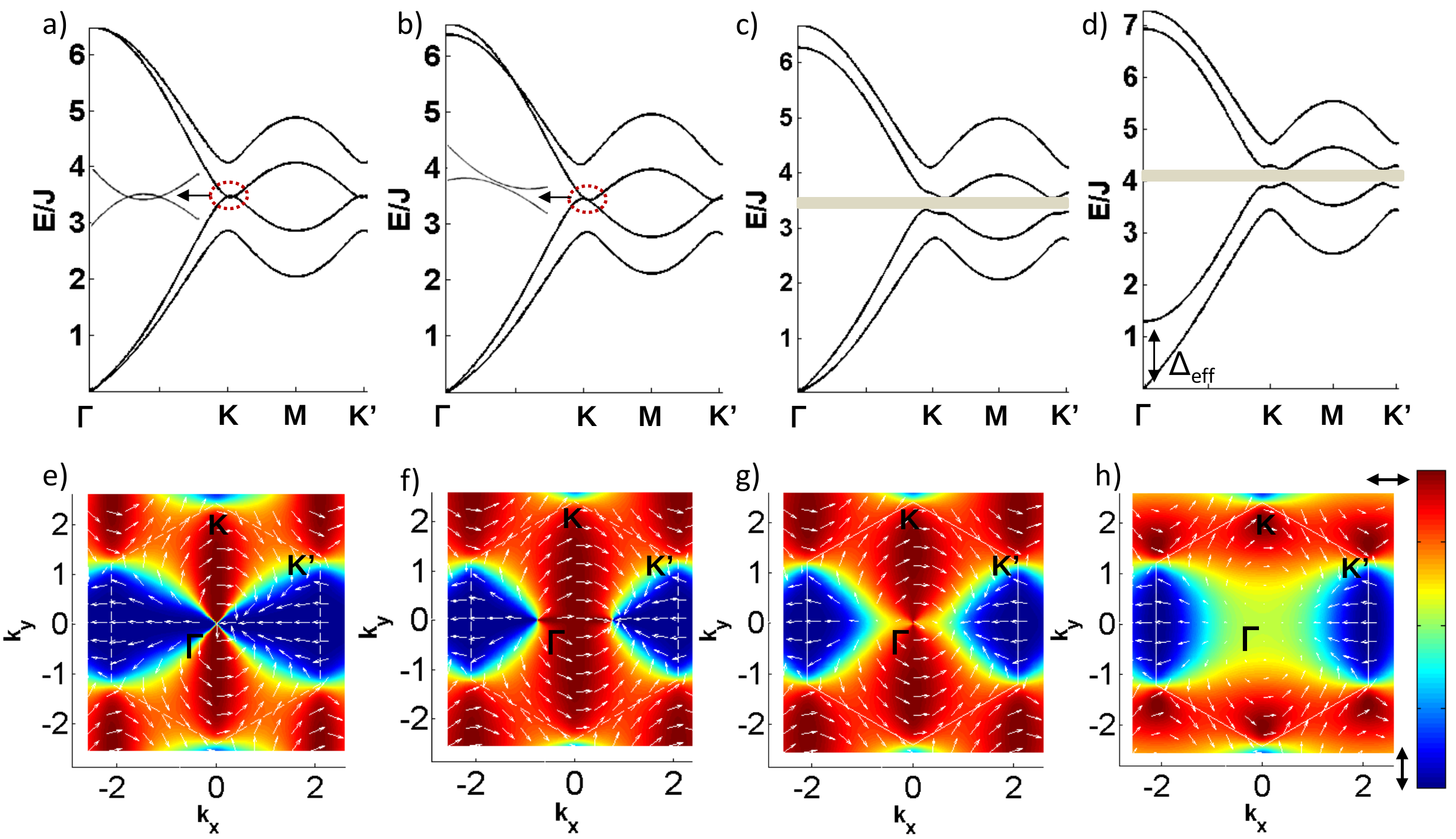}
 \centering
 \caption{ (Color online) Band structures and polarization textures of Bogoliubov excitations without and with applied magnetic field. (a)-(b) Energy dispersions without applied magnetic field: $\alpha_2=0,-0.2\alpha_1$ respectively. (c) Dispersions with applied magnetic field $B<B_C$, (d) with $B>B_C$. (e)-(f)-(g)-(h) Corresponding polarization and pseudospin textures. The colors represent the linear polarization degree while the white arrows show the corresponding pseudospin orientation of the lowest energy state.
 (Set of parameters: $\delta J=0.2J$, $\alpha_1n=J$)}
 \label{dispersionsetpolarisations}
 \end{figure}

 \end{widetext}

Because of the spin-anisotropic interactions the minimal energy configuration of a polariton condensate corresponds to a  linearly polarized state\cite{PhysRevLett.97.066402}. For the sake of definiteness, we choose at $k=0$: $\textbf{e}=(1,1)^T$ which gives a well defined (horizontal) orientation of the in-plane polarisation. In this configuration, the chemical potential is given by $\mu=(\alpha_1n+\alpha_2n)/2-3J$, where $n$ is the density, $J$ is  the tunneling coefficient defined above.
Using the Bogoliubov formalism, we compute the bogolon dispersion and the corresponding eigenstates. The results for the zero magnetic field case are shown in Fig. \ref{dispersionsetpolarisations}(a),(b),(e),(f).

First, we analyze the simpler case of $\alpha_2=0$, which is still spin-anisotropic since $\alpha_1\neq\alpha_2$. Panel (a) of Fig. \ref{dispersionsetpolarisations} shows the bogolon dispersion, and panel (e) -- the polarization texture: white arrows indicate the in-plane pseudospin direction and color shows the linear (H/V) polarization degree. Close to $\Gamma$, the two bogolon dispersions are TE and TM polarized. In other words, the TE-TM symmetry of the system is not broken by the presence of a condensate polarized along a well-defined spatial direction. The reason for this is that $\alpha_2$ is zero, as we shall see below.

 Due to interactions, close to $k=0$ these two dispersions depend linearly on the wave vector, with a speed of sound given by $c_{te,tm}=\sqrt{\alpha_1 n/m_{te,tm}}$, and the energy splitting (the magnitude of the TE-TM effective field) is also linear in $k$. At larger momenta, depending on the parameters, one may recover a parabolic dispersion with a parabolic splitting. Close to the K point, the preserved TE-TM symmetry of the system makes the dispersion and polarization structure similar with the ones obtained for polariton graphene with TE-TM splitting in the linear regime (see Fig. 2  of Ref.\cite{nalitov2015spin}): there is a linear intersection in K point, and 3 additional linear intersections appear because of the trigonal warping. The effective field texture around the K point is the one of the Dresselhaus field, as shall be discussed below.

The case with $\alpha_2\neq0$ is shown on the Figs.\ref{dispersionsetpolarisations}(b),(f). We have chosen to use $\alpha_2=-0.2\alpha_1$ for all calculations, although the exact value depends on the configuration (in particular, on the detuning). In general, $\alpha_2$ is usually negative \cite{Renucci2005} and small with respect to $\alpha_1$.
The scattering rates leading to the creation of a bogolon with polarisations parallel and perpendicular to the one of the condensate are therefore different, which affects the sound velocities of the dispersions at $k=0$. The eigenstates close to the $\Gamma$ point correspond to these two polarisations, with their slope given by $c_{\parallel}=\sqrt{(\alpha_1+\alpha_2)n/m^*}$ and $c_{\perp}=\sqrt{(\alpha_1-\alpha_2)n/2m^*}$ respectively (where $m^*$ is the reduced mass). The polarization texture is hence strongly modified in the center of the Brillouin zone, which is the consequence of the symmetry breaking by an in-plane field, see Fig.\ref{dispersionsetpolarisations}(f).

Due to this symmetry breaking, the form of the bands changes: the trigonal warping initially present around the K point, as in the linear case, disappears (\ref{dispersionsetpolarisations}(b). Two Dirac cones appear instead of four around each K point. Moreover, the branches of the same band cross each other, which modifies the polarization texture of each branches. We point here that this effective field does not open a gap, moreover, because the contribution of $\alpha_2$ changes the energy of the condensate, it can even lead to the closing of a gap, destroying the effect of the applied magnetic field. This will be discussed in details in further sections, but in general the effect of $\alpha_2$ is not the central subject of this work.

Indeed, in the following, we study the polariton graphene under an applied magnetic field, which overcomes the anisotropy induced by nonzero $\alpha_2$, and the effective in-plane field in K-K' valleys come back in its Dresselhaus-like form.

  \subsection{Nonzero magnetic field}
  As in the linear case, we apply an external magnetic field in order to open a gap and compute the topological invariants (Chern numbers) characterizing the band structure of bologons. In this section, we first remind the spin-Meissner effect, predicted for polariton condensates in planar cavities. Then, taking account this phenomenon, we come back to the study of bogolon dispersions in polaritonic graphene. In general, an applied magnetic field $B$ oriented in the growth direction leads to the polariton Zeeman splitting characterized by $-\Delta\sigma_z$ term in the Hamiltonian with $\Delta = |x|^2 g_X \mu_B B/2$, where $x$ is the exciton Hopfield coefficient, $g_X$ is the exciton $g$-factor, $\mu_B$ is the Bohr magneton.
  
  \subsubsection{The Spin-Meissner effect}
Before considering our specific problem of interest, it is certainly useful to remind the equilibrium behaviour of a 2D polariton condensate in the presence of a magnetic field. As shown in  \cite{Rubo2006}, at low magnetic field an equilibrium condensate in its ground state is elliptically polarized. It generates a self-induced Zeeman field which compensates the Zeeman splitting induced by the applied magnetic field. This is called the spin-Meissner effect, because of the analogy with the Meissner effect in superconductors, where an external magnetic field cannot penetrate into the superconducting region. Below a critical field $B_C$, the chemical potential is therefore constant against the magnetic field, the circular polarization degree of the condensate being given by:
 \begin{equation}
\rho_c=\frac{|\Psi^+|^2-|\Psi^-|^2}{|\Psi^+|^2+|\Psi^-|^2}
\end{equation}

Above the critical field, the condensate becomes circularly polarized. The splitting of the state with opposite circular polarization at $k=0$ is given by $\Delta_{eff}$.

\begin{equation}
\Delta_{eff}=\Delta-\Delta_C=|x|^2 g_X \mu_B (B-B_C)/2
\end{equation}
where the critical field $B_C$ is defined by the interaction energy via the corresponding splitting $\Delta_C$ :
\begin{eqnarray}
\Delta_C&=&\frac{(\alpha_1-\alpha_2)}{2}n\\
\Delta_C&=&|x|^2 g_X \mu_B B_C/2
\label{champCrit}
\end{eqnarray}
In the following, we are going consider the cases of subcritical ($B<B_C$) and supercritical ($B>B_C$) magnetic fields separately.

Another aspect, which will play a key role in the following, is that the energy renormalization induced by the interactions is larger for the bogolons states than for the condensate itself. Indeed, if one neglects $\alpha_2$, in the absence of the magnetic field the condensate is shifted by $\alpha_1 n/2$ with respect to the non-interacting case at $k=0$, whereas for large wavevectors the bogolon energy is shifted by $2\alpha_1 n/2$ with respect to the corresponding non-interacting dipsersion, as one can directly see from the Bogoliubov formula for the excitations $\hbar\omega=\sqrt{(\hbar^2 k^2/2m)^2+2\alpha_1 n/2 (\hbar^2 k^2/2m)}$, where the energy $\hbar\omega$ is measured from the chemical potential, itself being $\alpha_1 n/2$. The consequence is that the self-induced Zeeman field is larger for the excited states than for the ground state, while the Zeeman splitting induced by an applied field remains the same (unless the excitonic fraction changes significantly). Hence, in the magnetic field range $2B_C>B>B_C$, the bogolon dispersions with opposite circular polarisations should cross each other. The total Zeeman field (applied and self-induced) at large wave vector is in that case opposite to the one in the ground state.

  \subsubsection{Weak magnetic field $B<B_C$}
  
  We now turn back to the polariton graphene case considering an applied magnetic field weaker than the critical field $B_C$. 
  In such case, the condensate at equilibrium is elliptically polarized, with the polarization spinor being:
   \begin{equation}
  \textbf{e}=\begin{pmatrix}
  \cos\theta\\\sin\theta
  \end{pmatrix} ~,~~~ \theta=\frac{1}{2}\arcsin\sqrt{1-\left(\frac{B}{B_C}\right)^2}
\end{equation} 

By using this spinor in the expression of the condensate wavefunction $\mathbf{\Psi}$, we find the chemical potential: 
\begin{equation}
\mu= \frac{\alpha_1+\alpha_2}{2}n-3J
\end{equation}
Then we can compute the matrix for the Bogoliubov coefficients, that we can diagonalize in order to get the energy dispersions and polarisation texture (Fig. \ref{dispersionsetpolarisations}(c),(g)). In this section and the following one we restrict the consideration to the case of $\delta J < J/2$, typical for polariton graphene etched out of a planar cavity. The case of larger $\delta J$ is discussed in section III.

Figure \ref{dispersionsetpolarisations}(c) demonstrates a gap opening at K and K', similar to the linear case. Here, the self-induced Zeeman field completely compensates the external magnetic field around the $\Gamma$ point. The spin-Meissner effect is present, like in a planar microcavity without a periodic potential. The lower branches remain linear near the center of Brillouin zone.

The presence of the gap allows us to compute the band Chern numbers ($C_{C(V)}$) by using the gauge-invariant approach described above (Eq. \ref{chern}). In the weak field regime ($\Delta\ll B_C<\alpha_1 n/2$), the calculated Chern numbers $C_C=-2$, $C_V=+2$ are inverted with respect to the non-interacting case\cite{nalitov2015polariton}, where they have the values $C_C=+2$, $C_V=-2$, because the non-interacting case by definition corresponds to the opposite limit of large magnetic fields $\alpha_1 n=0 \ll \Delta$. This topological inversion occurs because of the self-induced Zeeman splitting at high wavevectors, which is opposite to the applied magnetic field and strongly exceeds it.
The system is in a different topological phase with inverted propagation direction of the chiral edge states, as compared with the linear case.

We stress here that if $\alpha_2$ becomes large (for example, in the regime of polariton Feshbach resonance \cite{Takemura2014}), the strength of the external field has to be large enough ($\Delta > \left|\alpha_2\right| n$) to overcome the anisotropy effect described in the previous section, in order to open a gap and obtain this topological insulating phase.

\subsubsection{Large magnetic field $B>B_C$}

  In this subsection, we consider the case of magnetic field above the critical value $B>B_C$. Hence, a Zeeman splitting appears between the two lower branches in the $\Gamma$ point, where interactions do not compensate the magnetic field anymore. The condensate polarization becomes circular in $\Gamma$: $\textbf{e}=(1,0)^T$
and the chemical potential becomes $\mu=\alpha_1 n -\Delta-3J$. We can observe that the chemical potential now depends of the applied magnetic field. Using the Bogoliubov approach as above, we find the eigenenergies and the eigenstates for this configuration. One of the two lower branches (polarized opposite to the condensate) becomes parabolic in $\Gamma$ and the degeneracy is lifted (see Fig. \ref{dispersionsetpolarisations}(d),(h)). The Zeeman splitting in the $\Gamma$ point is proportional to the difference between $B$ and $B_C$, exactly as in the spin-Meissner effect in planar cavities:
\begin{equation}
\Delta_{eff}=\Delta-\Delta_C=|x|^2 g_X \mu_B B_{eff}/2
\end{equation}

As we can see on Fig.\ref{dispersionsetpolarisations}(d), a gap opens between the valence and conduction bands. For $B$ just over $B_C$, the topology of the bands ($C_C=-2$, $C_V=+2$) is still inverted with respect to the non-interacting case. Since the gap does not close at $B_C$, the band topology cannot change, and this result is normal. However, when increasing the magnetic field again, the real Zeeman energy continues to rise whereas the self-induced Zeeman field, created by the interactions, is saturated. Consequently, there exists a threshold field $B=B_S$, where the net Zeeman field at K and K' cancels and the gap closes and immediately re-opens when $B>B_S$ with an inverted topology of the bands, associated with a change of the signs of the Chern numbers ($C_C=+2$, $C_V=-2$). This topology now corresponds to the non-interacting case. The phenomenon of the topology inversion is entirely due to the fact that interaction energy with the condensate is larger for large wavevector bogolons than in the center of Brillouin zone.\\

By writing the matrix in the upper $B_C$ case (appendix) in K point, we can find the analytical expression for the threshold magnetic field, at which the gap closes:
\begin{equation}
\Delta_S=\frac{1}{2}\left(\alpha_1n-\alpha_2n-3J \pm \sqrt{6J\alpha_1n+9J^2}\right)
\label{threshold}
\end{equation}
\begin{equation}
B_S=2\frac{\Delta_S}{|x|^2 g_X \mu_B}
\end{equation}

\section{Discussion}

\begin{widetext}

\begin{figure}[tbp]
\begin{center}
\includegraphics[scale=0.6]{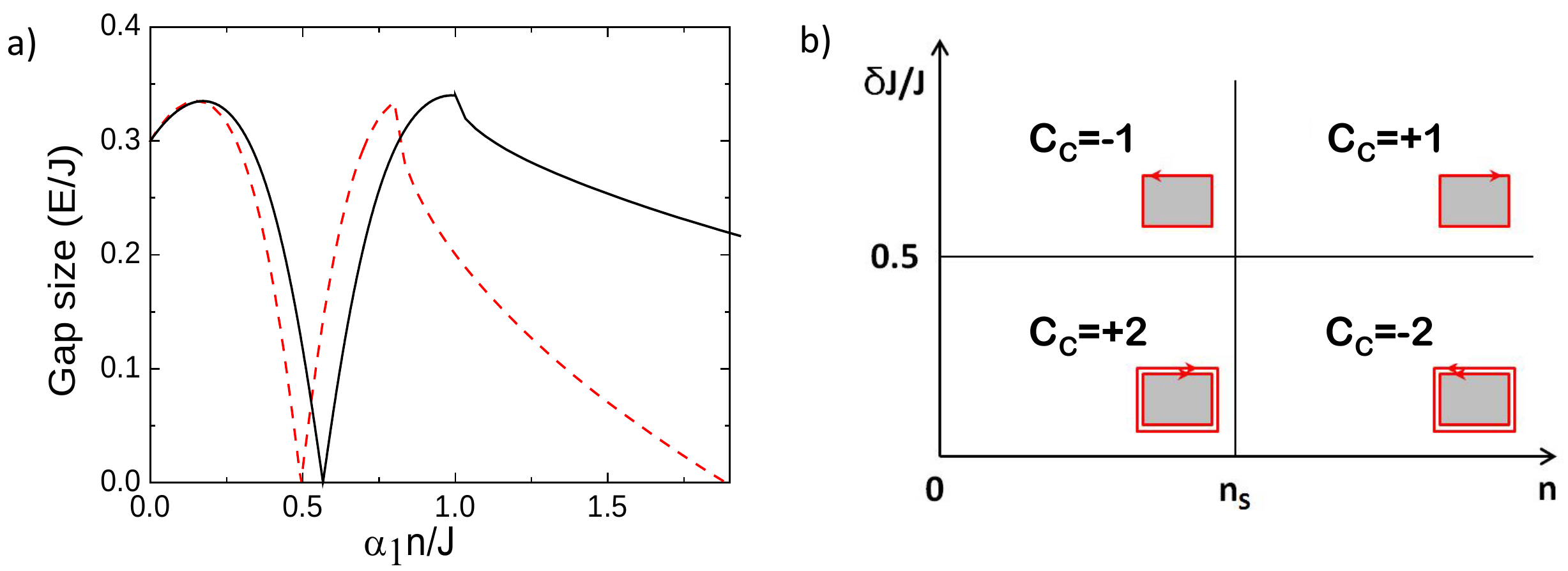}
\caption{ (Color online) Gap map and qualitative phase diagram. (a) Plot of the gap value under a constant magnetic field as a function of the density. The gap closes at the critical density $n=n_S$. (Parameters $\delta J=0.2J$, $\Delta=0.6 J$, black solid line: $\alpha_2=-0.2\alpha_1$, red dashed line: $\alpha_2=-0.5\alpha_1$). (b) The Chern numbers of the conduction band with a constant applied magnetic field. The insets illustrate the corresponding edge states.}
\label{phasescheme}
\end{center}
\end{figure}

\end{widetext}

In the previous section we have demonstrated that a topological inversion occurs at a threshold magnetic field $B_S$. The inversion of the topology at $B=B_S$ can be observed either varying the magnetic field or the polariton density. Whatever the density, if it is fixed, one can always increase the magnetic field to achieve the threshold value giving the inversion. If one fixes the magnetic field at some value, it is possible to observe the transition in the opposite direction: from the normal topology of the non-interacting system to inverted topology of the strongly-interacting system. Indeed, in polaritonic systems, the main experimentally adjustable parameter is the optical pumping, which controls the creation of polaritons, while their lifetime is generally fixed by the properties of the cavity. The evolution of the gap as a function of the condensate density is shown on the Fig.\ref{phasescheme}(a), where we can observe that the gap closes at a critical density $n_S$. The black solid line corresponds to the realistic value of $\alpha_2=-0.2\alpha_1$ (the same as in previous sections), while the red dashed line has been plotted with $\alpha_2=-0.5\alpha_1$, in order to clearly demonstrate the suppression of the topological insulator behavior by the anisotropy induced by the spin-anisotropic interactions of the condensate and its fixed linear polarization. Indeed, for any value of the applied magnetic field, one can always increase the density and close the gap via $\alpha_2 n$. Thus, the effect of this constant, which is usually relatively weak, can appear critical in certain configurations.

In the simplest configuration of relatively weak interactions, where $\alpha_1n<< J$, we can derive a simpler expression for the threshold magnetic field at which the inversion occurs, using the Taylor series expansion:
\begin{equation}
\Delta_S=\frac{2\alpha_1-\alpha_2}{2}n
\end{equation}
This result can be rewritten to give the threshold density as a function of an applied magnetic field:
\begin{equation}
n_S=\frac{2\Delta}{2\alpha_1-\alpha_2}~~,  ~~~ \Delta=|x|^2 g_x \mu_B \frac{B}{2}
\end{equation}

For $n<n_S$, the topology of Bogoliubov excitations is the same as in the linear case. The topology of the bare dispersion is transferred to the bogolon dispersion, as it has been already shown for atomic systems \cite{furukawa2015excitation}. Nevertheless, in the case of polaritons with their spin-anisotropic interaction and the resulting self-induced Larmor precession, the topology of the bogolon dispersion is inverted at the threshold density $n=n_S$ (Fig. \ref{phasescheme}). This result implies the inversion of the propagation direction of the topologically protected edge states.

Figure 6(b) demonstrates that the polariton graphene with an interacting quantum fluid can exhibit two topological transitions: one associated with the threshold density $n_S$ or magnetic field $B_S$, and the other associated with the relative strength of the spin-orbit coupling $\delta J/J=1/2$. At both transitions the signs of the Chern numbers are inverted, and so the propagation direction of the chiral edge states changes to the opposite one. The transition associated with the spin-orbit coupling leads also to the change of the magnitude of the Chern numbers and of the number of edge states: $C_{C,V}=\pm 2$ for $\delta J/J<1/2$ and $C_{C,V}=\mp 1$ for $\delta J/J>1/2$. We have also demonstrated that the gap closes at high interactions because of the effect of the singlet interaction constant $\alpha_2$. The presence of these two topological transitions makes the interacting polariton graphene an extremely rich system with a great potential for future experimental and theoretical studies. 

The potential of polariton graphene is confirmed by the coexistence of two dissipation-less phenomena, which are the superfluidity of the condensate and the existence of the topologically protected edge states. It will be especially useful to study the case of resonant pumping, which is the easiest feasible  experiment to study the macro-occupied  states in polaritonic systems. We can imagine interesting optical components, like the non-return valve for light, with its orientation controlled by a slight change of the particle density (external pumping). Our study is a part of a larger topic: the physics of bosonic quantum fluids in topological structures. A large variety of phenomena have already been studied for the physics of the condensates of cavity polaritons, like vortices, half-vortices, and solitons. These collective excitations, different from the bologons we study in the present work, are expected to give rise to new interesting phenomena in topologically non-trivial systems. 

\appendix*
\section{Example of the linearization for $B>B_C$}

The condensate is in the center of Brillouin zone and its polarization is completely circular: $\textbf{e}=(1,0)^T$. Therefore, we can write the solution for the weak excitations of the condensate as follows:

\begin{eqnarray}
\mathbf{\Phi}(\textbf{r},t)&=&(\Psi_A^+,  \Psi_A^-, \Psi_B^+, \Psi_B^-
)^T \nonumber \\ &=&\sqrt{n}\begin{pmatrix}1\\ 0\\1\\0  
\end{pmatrix}+\begin{pmatrix}u_A^+  \\ u_A^-\\u_B^+\\u_B^-
\end{pmatrix}e^{i(\textbf{k.r}-\omega t)}+\begin{pmatrix}v_A^+\\v_A^-\\v_B^+\\v_B^- 
\end{pmatrix}^*e^{-i(\textbf{k.r}-\omega t)} \nonumber\\ 
\end{eqnarray}
By inserting, this expression in the Gross-Pitaevskii equation, we find the expression for the chemical potential:

\begin{equation}
\mu=\alpha_1n-\Delta-3J
\end{equation}
Then we can deduce a system of 8 equations for the Bogoliubov coefficients $\mathbf{u},\mathbf{v}$, which can be written in matrix form :
\begin{widetext}
\begin{scriptsize}
\begin{equation}
M=\setlength\arraycolsep{0.1pt}\begin{pmatrix}
\alpha_1n+3J&0&-Jf_k&-\delta Jf_k^+& \alpha_1n&0&0&0\\
0&(\alpha_2n-\alpha_1n)+3J+2\Delta &-\delta Jf_k^-&-Jf_k&0&0&0&0\\
-Jf_k^*&-\delta J(f_k^-)^*&\alpha_1n+3J&0&0&0& \alpha_1n&0\\
-\delta J(f_k^+)^*&-Jf_k^*&0&(\alpha_2n-\alpha_1n)+3J+2\Delta&0&0&0&0\\
- \alpha_1n&0&0&0&-( \alpha_1n+3J)&0&Jf_{-k}^*&\delta J(f_{-k}^+)^*\\
0&0&0&0&0&(\alpha_1-\alpha_2)n-3J-2\Delta&\delta J(f_{-k}^-)^*&Jf_{-k}^*\\
0&0&- \alpha_1n&0&Jf_{-k}&\delta Jf_{-k}^-&-( \alpha_1n+3J)&0\\
0&0&0&0&\delta Jf_{-k}^+&Jf_{-k}&0&(\alpha_1-\alpha_2)n-3J-2\Delta
\end{pmatrix}
\end{equation}
\end{scriptsize}
\end{widetext}
Then we diagonalize numerically this matrix to obtain the eigenenergies and the eigenvectors.

\label{A1}

\bibliography{biblio2} 

\end{document}